\documentclass[%
 reprint,
 amsmath,amssymb,
 aps,
]{revtex4-2}

\usepackage{multirow}
\usepackage{verbatim}
\usepackage{graphicx}
\usepackage{dcolumn}
\usepackage{bm}

\begin{document}


\title{Dark Matter stimulated neutrinoless double beta decay.}

\author{Francesco Nozzoli}\email{Francesco.Nozzoli@cern.ch}
\author{Cinzia Cernetti}
\affiliation{%
INFN-TIFPA \& Phys. Dep. Trento University \\
 Via Sommarive 14 I-38123 Trento, It
}%

\date{\today}

\begin{abstract}
Nuclei that are unstable with respect to double beta decay are investigated in this work for a novel Dark Matter (DM) direct detection approach.  
In particular, the diagram responsible for the neutrinoless double beta decay will be considered for the possible detection technique of a Majorana
DM fermion inelastically scattering on a double beta unstable nucleus, stimulating its decay.
The exothermic nature of the stimulated double beta decay would allow the direct detection
also of a light DM fermion, a class of DM candidates that are difficult or impossible to investigate with the traditional elastic scattering techniques.
The expected signal distribution for different DM masses and the upper limits on the nucleus scattering cross sections, are shown and compared with the
existing data for the case of $^{136}$Xe nucleus. 
\end{abstract}

\maketitle


\section{Introduction}
Neutrinoless double beta decay is one of the pillars in the search for Physics beyond the Standard Model, this process allows to exploit the characteristics decay Q-value for a possible identification of the Majorana nature of the neutrinos and Leptonic quantum number violation \cite{Zeldovich,VALLE82}.
The Seesaw models provide a compelling mechanism to naturally generate the small neutrino mass (see e.g. \cite{Escudero_arxiv}) moreover, both the problem of baryon asymmetry in the Universe \cite{FUKUGITA1986} and different Dark Matter (DM)  candidates \cite{MANDAL2021,Arcadi_2020,DUTTAarxiv}, can be addressed by these models.

\begin{figure}[h]
    \centering
    \includegraphics[width = 0.49 \textwidth]{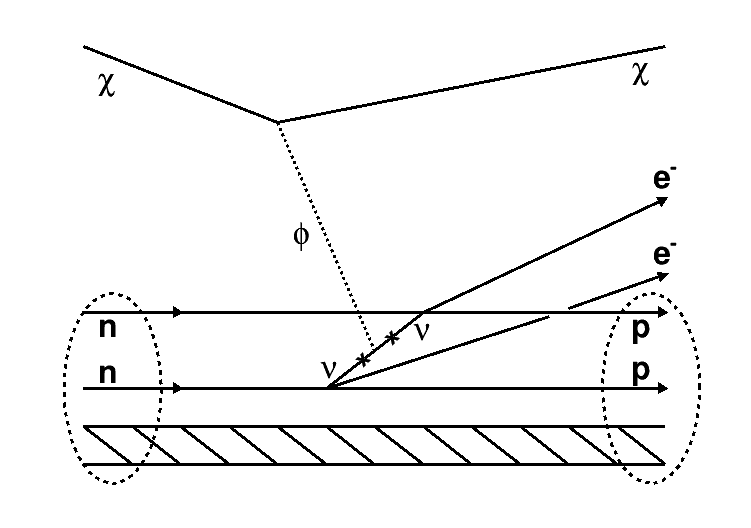}
    \caption{Example of a possible detection diagram for the Majorana DM fermion, $\chi$. The interaction through one or more Majoron fields $\phi$ stimulates a neutrinoless double beta decay of the nucleus (A,Z) to the daughter nucleus (A,Z+2). The kinetic energy of the two electrons is detected. A fraction of the decay Q-value is lost due to the upscattering $\chi$ particle, this would allow a possible measurement of the mass of $\chi$ particle from the energy distribution of the electrons.}
    \label{fig:feynmann}
\end{figure}

A typical phenomenology for these models is the addition of right-handed neutrino fields N$_R$ and a Majoron scalar field $\phi$, to the Standard Model (SM) Lagrangian:
\begin{multline}
\mathcal{L} = \mathcal{L_{SM}} + i\overline{N}_R \gamma^{\mu}\partial_{\mu}N_R + \left(\partial_{\mu}\phi \right)^{\dagger} \left( \partial^{\mu}\phi \right) -V(\phi) \\
-y_j\overline{l}^j_L H N_R - \frac{\lambda}{2} \overline{N}_R^c \phi N_R + h.c.
\end{multline}
where l$^j_L$=$\left( \substack{\nu_j \\ l_j^-} \right)$ are the SM lepton doublets ($j=e,\mu,\tau$) and H is the SM Higgs doublet.  
After spontaneous symmetry breaking of H and, possibly, of the $\phi$ scalar, the last two terms are providing a Dirac mass and a Majorana mass term, respectively. In the Seesaw mechanism, a very large value of the Majorana mass scale would naturally generate active neutrinos whose masses are much lighter than the (Dirac-) mass scale of the charged fermions.
Such a Lagrangian also provides two types of DM candidate: the Majoron that is a scalar particle \cite{Heeck_arxiv} and the sterile neutrino, the heavy mass eigenstate whose composition is dominated by the Majorana fermion N$_R$.
Focusing on the Majorana fermion DM candidates (in the following: $\chi$) the expected interactions (and self-interactions) are mediated by the Majoron, thus the direct detection of such a DM particle scattering on charged fermions could be very suppressed. 
On the other hand, the same diagram responsible for the possible neutrinoless double beta decay can be considered also for a possible detection technique of a Majorana fermion DM inelastically scattering on a double beta unstable nucleus, stimulating its decay, as shown in fig. \ref{fig:feynmann}.

In particular the exothermic nature of the stimulated double beta decay would allow the direct detection also of a light DM fermion, a class of DM candidates that are difficult/impossible to investigate with the traditional elastic scattering techniques\footnote{
The light dark matter detection is also possible through the conversion in a light neutrino or the beta decay of the nucleus \cite{DAMA_LDM,Robert1,Dror_2020,Robert2}.  
}.

It is important to note that the diagram shown in fig. \ref{fig:feynmann} is just one of the possible mechanisms for the DM particle to trigger a neutrinoless double beta decay of the nucleus. In this work we avoid focusing in the details of a specific interaction model, since beyond the sterile neutrino also other popular DM candidates are expected to be Majorana fermions (like, e.g., the supersymmetric Neutralino, Axino or Gravitino) 
thus they could interact the nucleus with similar phenomenology but different diagrams. 
Therefore we will focus on the expected signature for this novel detection technique and we will study the implication of current experimental results of neutrinoless double beta decay experiments.

\section{Expected energy distribution}
The available energy of a stimulated neutrinoless double beta decay of the nucleus (A,Z) to the daughter nucleus (A,Z+2) is provided by the reaction Q-value and by $\chi$ kinetic energy.
Due to the non relativistic nature of galactic DM particles, for simplicity we follow the approximation where the nucleus recoil
contributes to momentum conservation but provides a negligible contribution to the detected energy. Considering the MeV scale of Q-values for the typical nuclei adopted in double beta decay searches, these approximations holds with a reasonable accuracy both for light and heavy $\chi$ candidates.
Thus, the expected energy distribution can be evaluated following the Fermi's Golden Rule:
\begin{equation}
d\Gamma = \frac{|T_{fi}|^2}{4\pi^2\hbar}  \frac{d^3P_1}{(2\pi)^3}\frac{d^3P_2}{(2\pi)^3} \frac{d^3P_{\chi}}{(2\pi)^3} \delta(K_1+K_2+K_{\chi}-Q)
\end{equation}
where $K_{(1,2)}=\sqrt{P_{(1,2)}^2+m^2_e}-m_e$ are the kinetic energy of the electrons and 
$K_{\chi}=\sqrt{P_{\chi}^2+M^2_{\chi}}-M_{\chi}$ is the invisible kinetic energy carried away by the DM particle. 
The model dependent details of the DM particle interactions are encoded in the transition matrix $T_{fi}$, however, the overall behaviour of the expected distribution of the sum of electron's kinetic energy (that is experimentally detected) is shaped mainly by the phase space density factor (and also by detector effects).

\begin{figure}[h]
    \centering
    \includegraphics[width = 0.49 \textwidth]{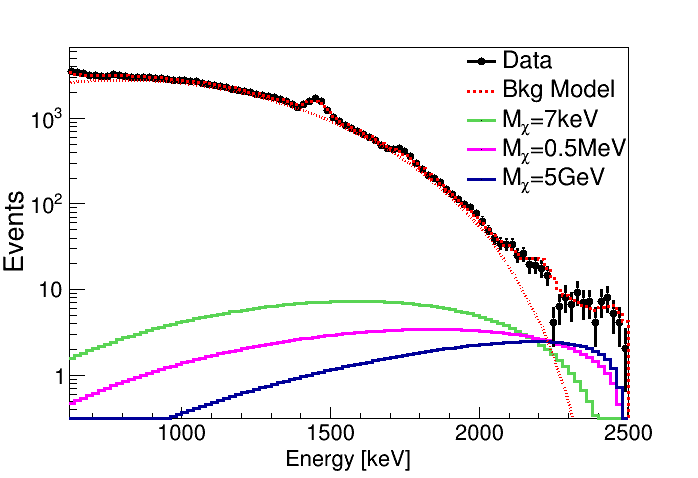}
    \caption{Example of upper limits on Dark Matter signals inferred analysing the 116.7 kg x day exposure collected by EXO-200 Phase-II (points).
    Maximum signals allowed at 90\% C.L. for M$_{\chi}$=7keV, 0.5MeV, 5GeV are shown as green, magenta and blue lines, respectively.
    }
    \label{fig:Exo_result}
\end{figure}

In the following, with the aim to study the phenomenology of this DM detection approach, the approximation of a constant matrix element is applied; the assumption of different interaction models can modify the details of the detected energy distribution but preserving the global mass/energy dependencies. 
In figure \ref{fig:Exo_result} the example of expected (sum-) energy distributions for the case of $^{136}$Xe target nuclei (Q=2.48MeV) is shown. In particular, the expected energy distributions are dependent on M$_{\chi}$; the case of M$_{\chi}\ll$m$_e$ (green line) provides a  distribution very similar to the one expected for the Majoron emitting neutrinoless double beta, $0\nu\beta\beta M$   (n=2) decay \cite{PhaseSpace,EXO_PRD}.    
 Increasing the value of M$_{\chi}$ an harder electron energy spectrum is expected. In particular, the case of M$_{\chi}\simeq $m$_e$ (magenta line) provides a distribution very similar to the one expected for the $0\nu\beta\beta M$ (n=1) decay, while the case M$_{\chi}\gg$m$_e$ provides a  much harder distribution (blue line).

Figure \ref{fig:Exo_result} also shows, as a comparison, the energy distribution measured by the EXO-200 experiment  (Phase-II 116.7 kg x yr, black points) \cite{EXO_PRD}. The published model of the detector background is also superimposed (red dashed line). The detector background is generally due to radioactive contaminants of the detector and surrounding materials, however, below the Q-value, it is dominated by the known $2\nu\beta\beta$ decay. It is important to note that the energy distribution of this last process (red dotted line) is very different as compared with the one expected by DM induced events, peaking at much lower energy.

No evidence for the $0\nu\beta\beta M$ decay was found so far in a dedicated study of the EXO-200 data \cite{EXO_PRD}, for the same reason, only upper limits to the DM induced neutrinoless double beta decay are shown in fig. \ref{fig:Exo_result}.

\begin{figure}[h]
    \centering
    \includegraphics[width = 0.49 \textwidth]{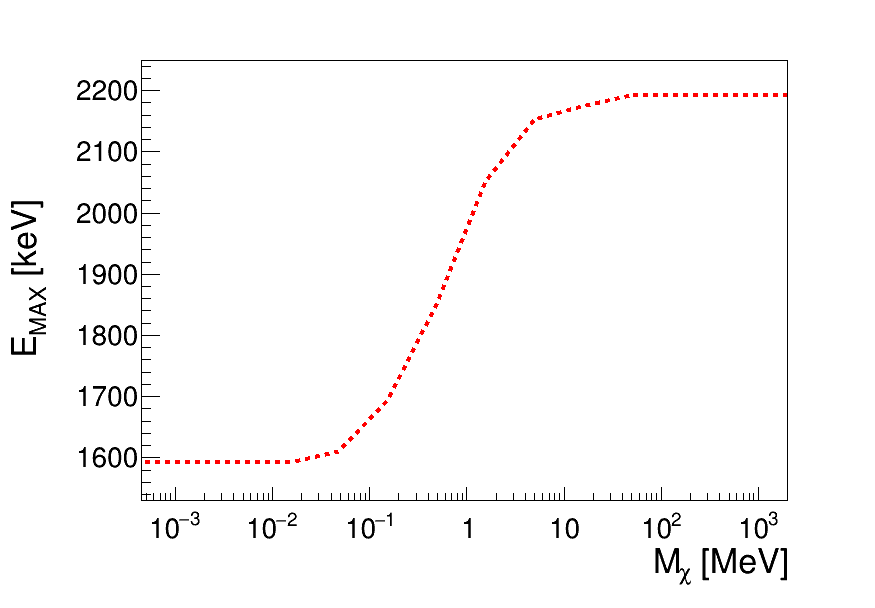}
    \caption{Expected behaviour of maximum of the detected energy distribution versus the Dark Matter particle mass inducing a neutrinoless double beta decay in $^{136}$Xe.    }
    \label{fig:Mass_xe}
\end{figure}
Finally, the expected behaviour of the maximum of the detected energy distribution in $^{136}$Xe as a function of $M_{\chi}$ is shown in fig. \ref{fig:Mass_xe}.
Thanks to the deformation of the electrons energy distribution caused by the upscattering of $\chi$ particle, a direct measurement of the Dark Matter particle mass could be feasible, in principle, for DM masses in the range 100keV-10MeV.

\section{Conclusion and Outlooks}
The possibility of a DM induced neutrinoless double beta decay is considered in this work, this could allow the investigation of light fermionic DM (like the 7.1 keV sterile neutrino \cite{sn7kev}) that is very difficult or impossible to detect with the elastic scattering technique.
For sub-MeV DM, the expected energy distribution for a DM induced decay, is similar to the expected distribution for $0\nu\beta\beta M$ decay (due to the large variations possible in Majoron and DM models) however some important differences among these rare processes can be noticed.
The first difference is based on the phenomenological point of view, in particular the $0\nu\beta\beta M$ decay is allowed only for relatively light Majorons (M$_{\phi}<$Q-value) while the DM induced decay could be mediated also by an heavy Majoron. 
On the other hand, a characteristics of the DM induced double beta decay is the expected annual modulation due to the yearly variation of the DM flux/velocity; this effect, in principle, could be exploited to disentangle the two different rare processes.

Finally, it is interesting to focus in the sensitivity of this approach for the detection of light dark matter. The direct detection of sub-GeV dark matter in the current underground experiments relies in the so called "Migdal effect", the DM induced atomic shake-off, pointed out by one of us in \cite{Dama_migdal}.  
\begin{figure}[h]
    \centering
    \includegraphics[width = 0.49 \textwidth]{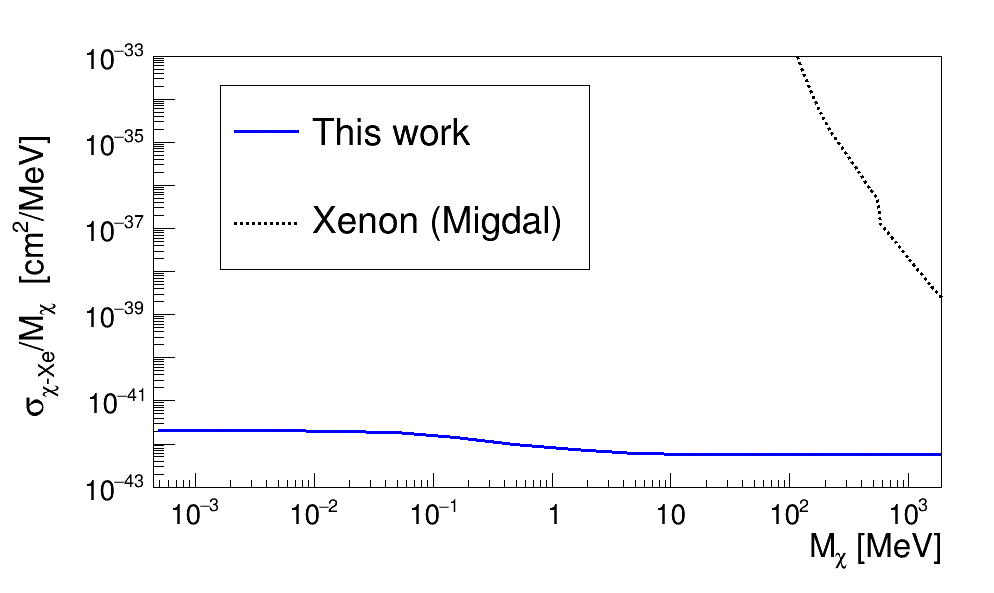}
    \caption{Example of upper limits on the total Dark Matter - Xenon nucleus cross section, obtained in this work considering the 116.7 kg x day exposure collected by EXO-200 Phase-II (blue line). Dotted line shows, as a comparison, the upper limits for the Dark Matter-Xenon nucleus elastic scattering for the very low mass region accessible exploiting the "Migdal effect" \cite{Dama_migdal,MigdalXe}.}
    \label{fig:Exo_Limits}
\end{figure}
In figure \ref{fig:Exo_Limits} the upper limits on the total nuclear scattering cross section ($\chi-^{136}$Xe) obtained in this analysis of EXO-200 Phase-II data are compared with the current upper limit on the DM-Xe nucleus scattering obtained by considering the "Migdal effect" in the XENON1T experiment \cite{MigdalXe}.
Despite a deeper comparison requires to detail the $\chi$-nucleus interaction model, the proposed approach could be very effective for the direct detection of light fermionic DM using the existing or future neutrinoless double beta decay experiments.

\nocite{*}

\bibliography{apssamp}

\end{document}